\documentclass[12pt]{article}
\usepackage{fancyhdr,amsmath,graphicx}
\newcommand{\mib}[1]{\mbox{\boldmath$#1$}}
\begin{document}
\begin{titlepage}
\vspace{5cm}
\begin{center}
{\LARGE ON-LINE LEARNING THROUGH SIMPLE PERCEPTRON LEARNING WITH A MARGIN}

\vspace{1cm}

{\Large Kazuyuki Hara$\dagger$}

\vspace{0.3cm}

{\large $\dagger$Department of Electronics and Information Engineering, \\Tokyo Metropolitan College of Technology}

\vspace{0.3cm}

{$\dagger$1-10-40 Higashi-oi, Shinagawa-ku, Tokyo 140-0011, Japan.\\Tel: 03-3471-6331, Fax: 03-3471-6338, E-mail: hara@tokyo-tmct.ac.jp}

\vspace{0.5cm}
 
{\Large Masato Okada$\ddagger$} 

\vspace{0.3cm}

{\large $\ddagger$Laboratory for Mathematical Neuroscience, \\Brain Science Institute, RIKEN} 

\vspace{0.3cm}

{$\ddagger$ 2-1 Hirosawa, Wako-shi, Saitama, 351-0198,Japan.\\Tel: 048-467-6845, Fax: 048-467-7349, E-mail: okada@brain.riken.go.jp}

\vspace{1cm}

{\bf\large Acknowledgement}
Part of this study has been supported by Grant-in-Aid for Scientific Research on Priority
Area No. 14084212 and Grant-in-Aid for Scientific Research (C) No. 13680472 and
 No. 14580438.

{\bf running head}

On-Line Learning with a margin
\end{center}

\end{titlepage}

\begin{abstract}
We analyze a learning method that uses a margin $\kappa$ {\it a la} Gardner
for simple perceptron learning. This method corresponds to the perceptron learning 
when $\kappa=0$, and to the Hebbian learning when $\kappa \rightarrow \infty$. 
Nevertheless, we found that the generalization ability of the method was superior to 
that of the perceptron and the Hebbian methods at an early stage of learning.
We analyzed the asymptotic property of the learning curve of this method through  
computer simulation and found that it was the same as for perceptron learning.
We also investigated an adaptive margin control method.
\end{abstract}

{\bf \large Keyword}

On-line learning, Margin, Simple perceptron, Generalization ability, Perceptron learning

\newpage

{\bf Symbols}

\begin{tabular}{ll}
$\mib{x}:$ & A input \\
$N$ : & Dimension of a input \\
$\mib{B}:$ & Teacher's weight vector \\
$\mib{J}:$ & Student's weight vector \\
$v:$ & Teacher's total input \\
$ul$ & Student's total input \\ 
$u:$ & Normalized student's total input \\
$\mbox{sgn}(\cdot)$ & Sign function \\
$\Theta(\cdot)$ & Threshold function \\
$\kappa$ :&Margin\\
$R:$ & Overlap between teacher and student weight vectors\\
$l:$ & Student's weight vector length \\
$\epsilon_g:$ & Generalization error \\
$\varphi :$ & Argument of teacher and student weight vectors\\
$P(u,v)$: & Distribution of $v$ and $u$ at limit of $N \rightarrow \infty$ 
\end{tabular}

\newpage
\section{Introduction}
\label{sec:intro}

Applying a margin to the decision boundary improves the generalization ability 
of many algorithms, for example, the support vector machine (SVM). 
Generalization ability is defined as the classification ability for 
learning samples not previously learned. The SVM places the decision boundary where the 
margin is maximized, and the support vectors are the learning samples closest to the 
decision boundary. 
In other words, the SVM improves the generalization ability by maximizing the margin 
as regards the learning samples already learned. 
Improving generalization ability is a key step towards solving the learning 
problem, and we believe that incorporating another form of learning - which we do by 
introducing the margin - can improve generalization ability.

Statistical mechanics has been used to analyze the learning ability of feed-forward neural 
networks or the effectiveness of information processing such as image restoration 
\cite{Nishimori1999}, and are often used to study the dynamics of learning or the 
ability of neural networks because they can depict the macroscopic dynamics of 
an object. 
The on-line learning of the simple perceptron, which consists of an input layer and 
an output unit, has been extensively studied using this approach \cite{Biehl1992}.
In the on-line learning, the network parameters are modified when a learning sample is 
presented and this sample is not used for feature learning. 

Perceptron learning \cite{Nishimori1999} $\sim$ \cite{Biehl1992} is a learning 
method applied through the simple perceptron. 
Learning occurs when the sign of the student's output differs from that of the teacher's 
output for an input. 

Requiring that the absolute value of the total input be larger than some margin, even if the 
sign of the student's output agrees with that of the teacher's output, would be 
beneficial because a learning sample of smaller absolute value of the total input than the 
margin is near the class boundary and can be easily moved to another classes by noise. 
Therefore, we propose a learning method in which a margin is applied {\it a la} Gardner 
\cite{Gardner1988} to a simple perceptron. 
Rosenblatt \cite{Rosenblatt1962} used similar learning algorithms.

Our algorithm is equivalent to perceptron learning when the margin is zero and equivalent 
to Hebbian learning when the margin is infinity. 
The method is thus intermediate between these two learning methods. 
We analyzed the dynamics of our algorithm through the statistical-mechanical method.
The dynamics of this learning method seems to be intermediate between those 
of perceptron learning and those of Hebbian learning.
Surprisingly, though, our learning algorithm is superior to the perceptron learning and the 
Hebbian learning in terms of the generalization error in the early stage of learning.

In Section 2, we review the theory of on-line learning, explain the generalization error, 
and give the order parameters we used to depict the learning dynamics.
In Section 3, we explain the formulation of our algorithm by showing the learning equation 
we use and deriving coupled macroscopic differential equations. 
These coupled differential equations are solved in Section 4, and the dynamics of our 
method are obtained.
The dependence of the generalization error on the margin in our algorithm is also discussed.
In Sections 5 and 6, respectively, we discuss the asymptotic property of our algorithm and the adaptive 
margin control method.

\section{THEORY OF ON-LINE LEARNING}

\subsection{Simple perceptron}

In this paper, we evaluate the learning ability by using the teacher-student formulation.
The teacher outputs the answer to the input.  
Learning ability is evaluated by how close the student's output is to the teacher's
output.
The teacher and the student are simple perceptrons and are formed with similar structures, 
as showed in Fig. \ref{network_structure}.
We assume that the input $\mib{x}$ is randomly selected according to a probabilistic distribution of 
$P(\mib{x})$.
Teacher outputs $\mbox{sgn}(v)$ correspond to the input $\mib{x}=\{ x_1, \ldots x_N \}$.

\begin{eqnarray}
\mbox{sgn}(v)&=& \mbox{sgn}\left (\sum_{i=1}^{N} B_i x_i \right ) = \mbox{sgn} \left (\mib{B} \cdot \mib{x} \right ) \\ 
\mib{B}&=&\{ B_1, \ldots B_N \}
\end{eqnarray}

\noindent
Here, $\mbox{sgn}(x)$ denotes the sign function that outputs $1$ when $x>0$ and outputs $-1$ when $x<0$.
The student outputs $\mbox{sgn}(ul)$ for the input $\mib{x}$ in the same way as the teacher.

\begin{eqnarray}
\mbox{sgn}(ul)&=& \mbox{sgn}\left (\sum_{i=1}^{N} J_i x_i \right ) = \mbox{sgn} \left (\mib{J} \cdot \mib{x} \right ) \\ 
\mib{J}&=&\{ J_1, \ldots J_N \}
\end{eqnarray}

\noindent
Here, $l$ is a proportional multiplier.

In the learning process, the student updates its weight vector according to the equation

\begin{equation}
\mib{J}^{m+1} = \mib{J}^m + f(v,u,l) \mib{x}.
\label{learning_equation}
\end{equation}

\noindent
Here, $m$ is the learning iteration number.
$f(v,u,l)$ is the function related to the learning algorithm used in the learning process.

\subsection{Assumptions}

There are two types of learning procedures - off-line learning and on-line
learning. 
In off-line learning, all the learning samples used in the learning are prepared beforehand.
The samples are fixed in each learning process, and the student iterates the updating using  
Eq. (\ref{learning_equation}). 
Here, the learning sample is the set of input $\mib{x}$ and its corresponding teacher's output.
In on-line learning, the student updates the weight vector by using a single sample.
The sample is not used again in the subsequent learning.
In this case, the input $\mib{x}$ and the student weight vector $\mib{J}$ become statistically 
independent when $N$ is sufficiently large, and the analysis becomes easy. 
Therefore, in this paper, we discuss learning ability based on on-line learning.

We consider the thermodynamic limit of $N \rightarrow \infty$ in the following discussions. 
The teacher's weight vector $\mib{B}$ is generated from random numbers taken from a Gaussian distribution 
of mean zero and unit variance. 
When $N$ is sufficiently large, the size of the weight vector becomes $|\mib{B}|=\sqrt{N}$. 
The student weight vector $\mib{J}$ is generated in the same way as the teacher's weight vector.
When the student weight vector $\mib{J}$ is updated, the size of the student weight
vector, denoted by $|\mib{J}|=l \sqrt{N}$, is changed, then we use the proportional 
multiplier $l$ and assume that the size of $l$ is finite.
The elements of the input vector are also generated from random numbers taken from a 
Gaussian distribution of mean zero and variance $1/N$; that is $|\mib{x}|=1$ when $N$ is 
sufficiently large. 
From the above formulation, self-averaging can be assumed. In the next paragraph, we briefly
explain this self-averaging. 

First, we consider the size of $\mib{J}$ and $\mib{x}$ to be of the same order.
When a input $\mib{x}$ is presented, whether the teacher and student outputs 
have the same sign is a statistical phenomenon. 
In other words, the student weight vector $\mib{J}$ will be updated to either 
$\mib{J} + \mib{x}$ or $\mib{J}$. 
Thus, there are $2^m$ possible states of $\mib{J}$ after the $m$-th learning iteration. 
However, $2^m$ statistical variables are too many to handle without difficulty. 
Hence, we assume that the absolute values of $\mib{J}$ and $\mib{B}$ are large compared with 
that of $\mib{x}$. 
Many learning samples would be required to increase the student weight vector 
length from $l$ to $l+dl$.   
To depict the trajectory of $l$, we need to consider only the statistical effects of the 
inputs. 
This treatment is called {\it self-averaging} in statistical mechanics.
In this manner, we introduced the above formulation to make the problem easier to handle.

There are two reasons for assuming the thermodynamic limit in the learning theory.
The first is that the deterministic differential equations of the order parameters 
$l$ and $R$ can be derived because the central limited theorem can be used at the 
thermodynamic limit.    
For example, the differential equation of $l$ is derived from Eq. (\ref{learning_equation});
however, $\mib{x}$ and $\mib{J}$ are random variables, so the equation becomes the 
random recurrence formula.  
Random variables $u=\mib{J} \cdot \mib{x}$ and $v=\mib{B} \cdot
\mib{x}$ follow the Gaussian distribution $P(u,v)$ of zero mean and 
unit variation when the input $\mib{x}$ is independent of the weight vector
$\mib{J}$, and the central limited theorem can be assumed.
Second, the generalization error is calculated by averaging
the error with respect to the input distribution $P(\mib{x})$. 

\begin{equation}
\epsilon_g = \int d\mib{x} P(\mib{x}) ( \mbox{sgn}(\mib{B}\cdot \mib{x}) - \mbox{sgn}(\mib{J} \cdot \mib{x}) )^2
\label{generalization_error1}
\end{equation}

\noindent
In general, this calculation is difficult because it requires the $N$-th multiple integral.
However, random variables $u$ and $v$ follow Gaussian distribution $P(u,v)$ because the central limited theorem can be used at the thermodynamic limit, so 
the generalization error can be calculated by averaging the error according to 
two-dimensional Gaussian distribution $P(u,v)$.

\begin{equation}
\epsilon_g = \int du dv P(u,v) ( \mbox{sgn}(v) - \mbox{sgn}(ul) )^2
\label{generalization_error2}
\end{equation}

\noindent
Moreover, the generalization error is calculated by using the direction cosine $R$
of $u$ and $v$ as shown by Eq. (\ref{generalization_error4}) in Sec. 2.4. 
As shown above, by assuming the thermodynamic limit, we can calculate the generalization error $\epsilon_g$.

\subsection{Conventional on-line learning algorithms}

In Hebbian learning, the weight vector $\mib{J}$ is updated using the equation

\begin{equation}
\mib{J}^{m+1}=\mib{J}^m+ \mbox{sgn}(v) \cdot \mib{x},
\label{hebb_rule}
\end{equation}

\noindent
where $m$ is the iteration number. 
Using Eq. (\ref{hebb_rule}), the weight vector $\mib{J}^m$ is updated according to the 
teacher output $\mbox{sgn}(v)$. 
The student input potential $u$ is not used to update the student weight vector $\mib{J}$.
In perceptron learning, the updating rule is 

\begin{equation}
\mib{J}^{m+1}=\mib{J}^m + \Theta (-uv)\cdot \mbox{sgn}(v) \cdot \mib{x}.
\label{perceptron_rule}
\end{equation}

\noindent
In this equation, the function $\Theta(x)$ returns 1 when $x>0$, and returns 0 when $x<0$. 
The use of this function in perceptron learning means that the weight vector is updated 
when the sign differs between the student's and the teacher's output.

\subsection{Relationship between generalization error and direction cosine}

The generalization error $\epsilon_g$ is used as a criterion for the quality of the learning.
In on-line learning, the generalization error is defined as the probability that a  
student who has learned $m$ learning samples will answer with an output different 
from the teacher's when the ($m+1$)-th input is presented. 
The overlap $R$ is one of the order parameters used to describe the dynamics of the 
generalization error. 
The overlap is the direction cosine of the weight vectors of the teacher and the student,  
which is defined as 

\begin{equation}
R = \frac{\mib{B}\cdot\mib{J}}{|\mib{B}||\mib{J}|} = \frac{1}{Nl}\sum_{j=1}^N B_j J_j .
\label{overlap}
\end{equation}

In Fig. \ref{overlap2}, the teacher weight vector $\mib{B}$ and the student weight 
vector $\mib{J}$ are depicted for an input dimension of $N=2$. 
The angle between $\mib{B}$ and $\mib{J}$ is denoted by $\varphi$. 
The input $\mib{x}$ is normalized as $|\mib{x}|=1$, then the inputs are 
distributed on the circumference of a circle with unit radius. 
The teacher output is $\mbox{sgn}(v)=\mbox{sgn}(\mib{B} \cdot \mib{x})$, so the input 
space is separated into two regions by a line orthogonal to the teacher weight 
vector $\mib{B}$. 
This line forms the class boundary. 
In the same way, the student output is $\mbox{sgn}(u)=\mbox{sgn}(\mib{J} \cdot \mib{x})$, 
so the input space is separated into two regions by a line orthogonal to the 
student weight vector $\mib{J}$. 
This line forms the decision boundary. 

Since $\varphi$ is the angle between the teacher weight vector $\mib{B}$ and the student 
weight vector $\mib{J}$, the teacher and the student outputs differ in the areas defined by 
the thick arcs along the unit circle in Fig. \ref{overlap2}. 
We assume that the input $\mib{x}$ is selected at random from the circumference of 
the unit circle, so the generalization error $\epsilon_g$ is given by the ratio of 
the circumference of the unit circle to the length of the thick arcs:

\begin{equation}
\epsilon_g=\frac{\varphi}{\pi}.
\label{generalization_error3}
\end{equation}

\noindent
$\varphi$ can be calculated as 

\begin{equation}
\varphi=\tan^{-1} \left( \frac{\sqrt{1-R^2}}{R} \right).
\label{angle}
\end{equation}

\noindent
From Eqs. (\ref{generalization_error3}) and (\ref{angle}), the generalization error 
is defined as 

\begin{equation}
\epsilon_g=\frac{\varphi}{\pi}=\frac{1}{\pi} \tan^{-1} \left( \frac{\sqrt{1-R^2}}{R} \right).
\label{generalization_error4}
\end{equation}

\noindent
Therefore, we can calculate the generalization error by using the overlap $R$ instead of Eq. (\ref{generalization_error2}).

Another aspect of the relationship between the generalization error $\epsilon_g$ and 
the overlap $R$ is as follows. 
As explained, the generated input $\mib{x}$ is statistically independent of the 
teacher weight vector $\mib{B}$. 
The student weight vector $\mib{J}$ and the new input $\mib{x}$ are also 
statistically independent. 
The distribution of the total input of the teacher $v$ and the normalized total input 
of the student $u$ therefore becomes a Gaussian distribution with mean zero, 
unit variance, and correlation of $R$. 
This distribution is denoted $P(u,v)$ and is written as

\begin{equation}
P(u,v)=\frac{1}{2 \pi \sqrt{1-R^2}} \exp \left ( - \frac{u^2+v^2-2Ruv}{2(1-R^2)} \right ), 
\label{distribution_uv}
\end{equation}

\noindent
where $R$ is the overlap defined by Eq. (\ref{overlap}).

Figure \ref{generalization_and_overlap}(a) depicts $P(u,v)$ when the overlap $R$ 
equals zero. 
The abscissa axis is the total input of the teacher and the ordinate axis is the 
total input of the student. 
Because $R=0$, from Eq. (\ref{distribution_uv}), the distribution forms a circle as 
shown in Fig. \ref{generalization_and_overlap}(a). 
Since the signs of $v$ and $u$ differ, an error occurs in the region marked by 
the oblique lines. 
On the other hand, as the learning progresses and overlap $R$ approaches a value of one, 
the distribution $P(u,v)$ asymptotically approaches the line of $v-u=0$. 
This is depicted in Fig. \ref{generalization_and_overlap} (b). 
Again, an error occurs in the area marked by the oblique lines, but this area is much 
smaller in Fig. \ref{generalization_and_overlap} (b) than in 
Fig. \ref{generalization_and_overlap}(a). 
This confirms that as overlap $R$ approaches one, the generalization error approaches zero.

\section{FORMULATION}
\label{sec:formulation}

\subsection{Learning equations of proposed algorithm}

The learning equation discussed in this paper is 

\begin{small}
\begin{eqnarray}
&&\mib{J}^{m+1}=\mib{J}^m + \nonumber \\
&&\Theta \left( \kappa - \left (\sum_j J_j x_j \right) \mbox{sgn}\left( \sum_j B_j x_j \right ) \right ) \mbox{sgn}(v) \mib{x}, 
\label{kappa_rule0}
\end{eqnarray}
\end{small}

\noindent
where $m$ is the number of iterations, and $\kappa$ is the margin.
In perceptron learning, if $(\sum_j J_j x_j) \mbox{sgn}(\sum_j B_j x_j)>0$, no 
learning occurs because the signs of the teacher and the student outputs are the same. 
However, it is a good idea to require that the size of 
$(\sum_j J_j x_j) \mbox{sgn}(\sum_j B_j x_j)$ be larger than the margin even if the signs 
of $v$ and $u$ are the same because a input of smaller total input than the 
margin will be near the class boundary and can be easily moved to another class by noise.
To do this, we introduced the margin $\kappa$ into perceptron learning as shown in 
Eq. (\ref{kappa_rule0}).
By rewriting Eq. (\ref{kappa_rule0}), we derived Eqs. (\ref{general_learning_equation}) and
(\ref{kappa_rule}): 

\begin{eqnarray}
\mib{J}^{m+1}&=&\mib{J}^m+f(v,u,l) \mib{x} \label{general_learning_equation}\\
f(v,u,l)&=&\Theta(- (l \mbox{sgn}(v) u - \kappa)) \mbox{sgn}(v) \label{kappa_rule}
\end{eqnarray}

We will explain the effectiveness of the margin $\kappa$ by using Fig. \ref{kappa}, which shows
the distribution of $P(u,v)$ when the overlap $R=0$.
In this figure, the abscissa axis is the total input of the teacher and the ordinate
axis is the total input of the student. 
In the region marked by the oblique lines, the signs of $v$ and $u$ differ.
Learning occurs in this region when perceptron learning is used, but does not occur in the 
other regions. 
When Hebbian learning is used, all the regions are the object of learning. 
The dashed lines depict the margin $\kappa$. 
Our algorithm enables learning when the absolute value of the total input of the student 
$|ul|$ is below the dashed line. 

As shown in Fig. \ref{kappa}, when $\kappa=0$, our algorithm enables learning within the 
regions marked by the oblique lines - the same learning region as for perceptron learning. 
This can be shown by applying $\kappa=0$ to Eq. (\ref{kappa_rule}): 

\begin{eqnarray}
f(v,u,l) &=& \Theta(-lu \mbox{sgn}(v)) \mbox{sgn}(v) \nonumber \\
         &=& \Theta(-uv) \mbox{sgn}(v)
\end{eqnarray}

On the other hand, as Fig. \ref{kappa} shows, when $\kappa$ is infinity, our method 
enables learning in all the regions, as is the case with Hebbian learning. 
And when $\Theta(\infty)$ equals 1, Eq. (\ref{kappa_rule}) can be rewritten as 

\begin{equation}
f(v,u,l)= \mbox{sgn}(v), 
\end{equation}
 
\noindent
which shows that our algorithm is equivalent to Hebbian learning when $\kappa$ is infinity.
Thus, our algorithm represents an intermediate form between perceptron and Hebbian learning.

When $0 < \kappa < \infty$, the learning occurs as follows.
From Eq. (\ref{kappa_rule}), our algorithm learns by using Hebbian learning when 
the absolute value of the total input $|ul|$ is below the margin $\kappa$ even if 
the signs of the teacher's and the student's output are the same.
Thus, our algorithm enables learning in regions where perceptron learning is not possible.
We therefore expect this method to be capable of generalization ability better than that of
perceptron learning. 

\subsection{Differential equations of learning dynamics}

Next, we will derive and analyze the coupled differential equations of the overlap $R$ 
and the length of student weight vector $l$. 
The overlap $R$ is the direction cosine of the teacher weight vector $\mib{B}$ and the 
student weight vector $\mib{J}$. 

As discussed in Sec. 2.2, we formulated that the size of the weight vector $|\mib{J}|$ is
$O(\sqrt{N})$ and that the size of the input vector $|\mib{x}|$ is 1. This means that
we need $N$ input vectors to have $\Delta \mib{J}$ changes.
Consequently, we define the learning iteration $m$ as $m=Nt$ and use the continuous variable $t$
to represent the learning process. 

By using this formulation, a time-dependent differential equation of the student weight vector 
length $l$ can be derived.
To obtain this differential, we square both sides of Eq. (\ref{general_learning_equation}).
By averaging the terms of Eq. (\ref{general_learning_equation}) by the distribution of 
$P(u,v)$, we obtain the differential equation for $l$. A more explicit derivation is given 
in the Appendix. 

The differential equation of the direction cosine $R$ is obtained by calculating the 
product of $\mib{B}$ and Eq. (\ref{general_learning_equation}). 
The differential equation of $R$ is then obtained through a calculation similar to that 
used for $l$. An explicit derivation is again given in the Appendix. 
The obtained coupled differential equations are 

\begin{eqnarray}
 \frac{dl}{dt} &=& \langle f u \rangle + \frac{\langle f^2 \rangle}{2l}, 
  \label{l^m_macro_dynamics_3} \\
 \frac{dR}{dt} &=& \frac{\langle f v \rangle - \langle f u\rangle R}{l} -
 \frac{R}{2 l^2} \langle f^2 \rangle .
  \label{R^m_macro_dynamics_3} 
\end{eqnarray}

\noindent
The symbol $\langle \cdots \rangle $ means averaging over the distribution $P(u,v)$.

The main purpose of the on-line learning theory described in this paper is to calculate
the generalization error $\epsilon_g$. We can calculate $\epsilon_g$ by using the order
parameters $R$ and $l$. Thus, we should know the time dependence of the order parameters
$R$ and $l$ to obtain that of the generalization error. The time dependence of the order
parameters $R$ and $l$ is described by Eqs. (\ref{l^m_macro_dynamics_3}) and 
(\ref{R^m_macro_dynamics_3}). To calculate Eqs. (\ref{l^m_macro_dynamics_3}) and 
(\ref{R^m_macro_dynamics_3}), we should know the statistical average of $\langle fv \rangle$, 
$\langle fu \rangle$ and $\langle f^2 \rangle$ with respect to $v=\mib{B} \cdot \mib{x}$ and  
$ul=\mib{J} \cdot \mib{x}$. Then $\langle fu \rangle, \langle fv \rangle$, and $\langle f^2 \rangle$ 
are calculated for our algorithm. 

$\langle fu \rangle$ is calculated by averaging the product of Eq. (\ref{kappa_rule}) 
and the total input of the teacher $v$ over $P(u,v)$.
$\langle fv \rangle$ is calculated by averaging the product of Eq. (\ref{kappa_rule})
and the total input of the student over $P(u,v)$. 
$\langle f^2 \rangle$ is calculated by averaging the square of Eq. (\ref{kappa_rule})
over $P(u,v)$.
The results are shown in Eqs. (\ref{fu_kappa_algorithm}), (\ref{fv_kappa_algorithm}), 
and (\ref{f2_kappa_algorithm}).

\begin{eqnarray}
\langle fu \rangle &=& \int du dv P(u,v) f(v,u,l) u \nonumber \\
     &=& \frac{2R}{\sqrt{2 \pi}} 
H \left( \frac{- \frac{\kappa}{l}}{\sqrt{1-R^2}} \right) - \nonumber \\
&& \sqrt{\frac{2}{\pi}} \exp \left( - \frac{\kappa^2}{2l^2} \right)  
  H \left( \frac{- \frac{\kappa}{l}R}{\sqrt{1-R^2}} \right) 
\label{fu_kappa_algorithm}
\end{eqnarray}
\begin{eqnarray}
\langle fv \rangle &=& \int du dv P(u,v) f(v,u,l) v \nonumber \\
     &=& \sqrt{\frac{2}{\pi}}
 H \left( \frac{- \frac{\kappa}{l}}{\sqrt{1-R^2}} \right) - \nonumber \\
&& \frac{2R}{\sqrt{2\pi}}  \exp \left( - \frac{\kappa^2}{2l^2} \right) 
H \left( \frac{- \frac{\kappa}{l}R}{\sqrt{1-R^2}} \right)
\label{fv_kappa_algorithm}
\end{eqnarray}
\begin{eqnarray}
\langle f^2 \rangle &=& \int du dv P(u,v) f^2(u,v,l) \nonumber \\ 
  &=& 2 \int_0^{\infty} Dv
  H \left( \frac{Rv-\frac{\kappa}{l}}{\sqrt{1-R^2}} \right)
\label{f2_kappa_algorithm}
\end{eqnarray}

\noindent
where,

\begin{eqnarray}
Dx&=&\frac{dx}{\sqrt{2 \pi}}\exp \left(-\frac{x^2}{2} \right) \\
H(u)&=&\int_u^{\infty}Dx.
\end{eqnarray}

We substitute zero for $\kappa$ in Eqs. (\ref{fu_kappa_algorithm}) $\sim$ 
(\ref{f2_kappa_algorithm}) and are identical with $\langle fu \rangle, \langle fv \rangle$, 
and $\langle f^2 \rangle$ of perceptron learning.  
Likewise, we substitute infinity for $\kappa$ in Eqs.  (\ref{fu_kappa_algorithm}) $\sim$ 
(\ref{f2_kappa_algorithm}) and are identical with those of Hebbian learning.

\section{Results}

  First, we will consider the results for $\kappa=10$.
The generalization error was calculated analytically by applying the overlap $R$ to Eq. 
(\ref{generalization_error4}). 
$R$ was obtained by numerically solving Eqs. (\ref{l^m_macro_dynamics_3}) and 
(\ref{R^m_macro_dynamics_3}).
The generalization error curve obtained through the analytical calculation is shown in 
Fig. \ref{generalization_error_kappa1}(a). 
In this figure, analytical results (solid line labeled by ``ana'') and a numerical 
simulation results (dashed line labeled by ``num'') are shown. The input dimension $N$ used 
in the numerical simulation was 1000. 
These are almost identical, and the numerical simulation results
are distributed around the analytical results. 
The time step corresponded to the presentation of $N$ learning samples.
The results for $\kappa=0$, which corresponds to perceptron learning, 
and for $\kappa \rightarrow \infty$, which corresponds to Hebbian learning, 
are also shown in Fig. \ref{generalization_error_kappa1}(a) for comparison. 
Because our algorithm represents an intermediate form between perceptron and 
Hebbian learning, we expected the learning dynamics of Eq. (\ref{kappa_rule}) to be 
midway between those of perceptron learning and those of Hebbian learning. 
However, the generalization error of our algorithm was lower than that of either 
alternative learning method from $t=10$ to $200$. 
Therefore, at an early stage of learning, our algorithm seems to be superior to both 
perceptron and Hebbian learning in this respect.
The generalization error for each time step was calculated using Eqs. 
(\ref{general_learning_equation}) and (\ref{kappa_rule}). 

Next, we analyzed how the margin $\kappa$ affected the generalization error.
Figures \ref{generalization_error_kappa1}(b) and \ref{generalization_error_kappa2}(a) and 
(b) show the results for $\kappa$ of 1, 100, and 1000. These figures show results of the 
analytical calculation and the numerical simulation.  
As shown, the generalization error with our algorithm tended to be lower than that of both 
perceptron learning and Hebbian learning, particularly at the early stage of learning, 
for every margin.
Therefore, we expect similar behavior with any margin $\kappa$.

Moreover, we analyzed for the margins $\kappa=10^{-5}$ and $\kappa=10^5$. 
As we explained in Sec. 3.1, when the margin is relatively small, our algorithm will 
performe as perceptron learning, and when the margin is relatively large, it will 
performe as Hebbian learning. 
In Fig. \ref{generalization_error_kappa3}(a) and (b), the analytical 
results for $\kappa=10^{-5}$ and $\kappa=10^5$ are shown. 
The results for perceptron learning and Hebbain learning are also shown for comparison. 
In Fig. {\ref{generalization_error_kappa3}(a), the results for $\kappa=10^{-5}$ (solid 
line labeled by ``10\verb+^{-5}+(ana)'') are plotted until $t=7500$ to show how the results 
matched to the perceptron's (dashed line labeled by ``Perceptron''). The results 
for $\kappa=10^{-5}$ and perceptron are identical. 
In Fig. {\ref{generalization_error_kappa3}(b), the results for $\kappa=10^5$ was identical
to the results for Hebbian learning. 

\section{Asymptotic property}

The asymptotic property of the generalization error of perceptron learning is known to be 
$t$ to the power of $-1/3$ and the asymptotic property of the generalization ability of Hebbian 
learning is $t$ to the power of $-1/2$ \cite{Nishimori1999}. 
Therefore, we investigated the asymptotic property of the generalization error of our 
method in the region of $t > 10000$. 
The results are shown in Table \ref{asymptotic_property_kappa}.

The asymptotic property with our algorithm was close to $t$ to the power of $-1/3$. 
In the region of large $t$, the length of the student weight vector $l$ increased
monotonically. 
In this case, the margin $\kappa$ can be considered constant along time $t$, so at the 
limit of $t \rightarrow \infty$, the actual margin $\kappa/l$ will converge to zero. 
The asymptotic with our algorithm will then equal that with perceptron learning. 
Thus, we consider the asymptotic property with our algorithm to be $t$ to the power of $-1/3$.  

\section{Adaptive margin control}

The generalization ability of our method became superior to that of Hebbian and perceptron 
learning when we introduced the margin into perceptron learning. 
This occurred when $t$ was close to the margin $\kappa$, and our method converged toward 
the dynamics of perceptron learning. 
Thus, we naturally think that adjusting the margin with respect to the learning time might make 
our method superior to both Hebbian and perceptron learning.
For instance, this could be done by setting the margin to some small value in the early 
stage of learning, and gradually enlarging it. 
We tried to find the optimum value of $\alpha$ for $\kappa=\alpha l$ to overcome the 
generalization error of Hebbian learning.

In Fig. \ref{Adaptive_margin1}(a), the margin was controlled so that $\kappa=l$.
The generalization ability of our method was improved and was superior to that of Hebbian learning
when $1<t<100$.  
However, controlling the margin failed for $t>100$. 
Controlling the margin fully succeeded, though, when we adjusted the margin $\kappa$ to $1.5l$
(Fig. \ref{Adaptive_margin1}(b)).
In this case, however, the difference in the generalization errors of our method and Hebbian 
learning was small.
We also investigated the case where $\kappa=2l$ (Fig.\ref{Adaptive_margin2}). 
The generalization error of our method was smaller than that of Hebbian learning, but 
the difference between the two methods was smaller than that of Fig.\ref{Adaptive_margin1}(b). 

\section{CONCLUSIONS}

We have described a new learning method that uses the margin $\kappa$ {\it a la} Gardner 
for perceptron learning. 
This method can correspond to either Hebbian learning or perceptron learning 
depending on the size of $\kappa$. 
Coupled differential equations of order parameters $R$ and $l$, where $R$ is the overlap of 
the teacher weight vector $\mib{B}$ and the student weight vector $\mib{J}$, and $l$ is the 
length of the student weight vector, have been derived for our algorithm. 
Our analytical results show that the generalization error with our algorithm tends to be 
lower than that of either Hebbian or perceptron learning at the early stage of learning 
over a wide range of $\kappa$. 
Also, the asymptotic property of the generalization ability with our algorithm was equal 
to that of perceptron learning. 
Moreover, we investigated the effect of margin adaptation and found that the generalization 
error of our method was superior to that of Hebbian learning when we adjusted the margin to 
$\kappa=1.5l$. 
However, the improvement in the generalization error was small.
In our future work, we plan to compare our method with other learning methods that use 
an adaptive learning coefficient \cite{Kinouchi1992}.

\noindent
{\bf\large Acknowledgement}

Part of this study has been supported by Grant-in-Aid for Scientific Research on Priority
Area No. 14084212 and Grant-in-Aid for Scientific Research (C) No. 13680472 and No. 14580438.

\begin{small}

\end{small}

\appendix

\section{Derivation of differential equations of order parameter $R$ and $l$}

First, we derive Eq. (\ref{l^m_macro_dynamics_3}). 
We square both sides of Eq. (\ref{general_learning_equation}).
For simplicity, we denote $f(v,u,l)$ as $f$.

\vspace{-0.5cm}

\begin{eqnarray}
\mib{J}^{m+1} \cdot \mib{J}^{m+1} &=& \mib{J}^m \cdot \mib{J}^m + f^2 \mib{x} \cdot \mib{x} \nonumber \\
&+& 2f\mib{J}^m \cdot \mib{x}
\label{l^m_macro_dynamics_4}
\end{eqnarray}

\noindent
From $|\mib{J}^m|=l^m \sqrt{N}$ and $u=\mib{J}\cdot \mib{x}$, Eq. (\ref{l^m_macro_dynamics_4}) becomes 

\begin{equation}
N(l^{m+1})^2 = N(l^m)^2 + f^2 + 2 f l^m u_m 
\label{l^m_macro_dynamics_5}
\end{equation}

\noindent
Averaging Eq. (\ref{l^m_macro_dynamics_5}) over the distribution $P(u,v)$ of the teacher's total 
input $v$ and the normalized student's total input $u$ and assuming self-averaging for $l$, 
we rewrite Eq. (\ref{l^m_macro_dynamics_5}) as the next equation. Here, averaging is denoted as 
$\langle \cdots \rangle$. 

\begin{equation}
N(l^{m+1})^2 = N(l^m)^2 + \langle f^2 \rangle + 2l^m \langle fu_m \rangle .
\label{l^m_macro_dynamics_6}
\end{equation}

\noindent
At the thermodynamic limit, Eq. (\ref{l^m_macro_dynamics_6}) becomes a differential equation. 
Equation (\ref{l^m_macro_dynamics_6}) is rewritten as, 

\begin{equation}
N(l^{m+1}+l^m)(l^{m+1}-l^m) = \langle f^2 \rangle + 2l^m \langle fu_m \rangle .
\end{equation}

\noindent
We substitute $l^m = l$, $l^{m+1} = l+dl$, $u_m=u$ and $1/N \rightarrow dt$, and then simplify the equation.
The next equation is then given and Eq. (\ref{l^m_macro_dynamics_3}) is derived.

\begin{equation}
\frac{dl}{dt} = \langle fu \rangle + \frac{\langle f^2 \rangle}{2l} 
\label{l^m_macro_dynamics_7}
\end{equation}

To derive the differential equation for $R$, we multiply both sides of Eq. 
(\ref{general_learning_equation}) by $\mib{B}$ to obtain

\begin{eqnarray}
\mib{B} \cdot \mib{J}^{m+1} &=& \mib{B} \cdot \mib{J}^m  + f \mib{B} \cdot \mib{x} \nonumber\\
Nl^{m+1}R^{m+1} &=& Nl^mR^m + f v .
\label{R^m_macro_dynamics_4}
\end{eqnarray}

\noindent
Equation (\ref{R^m_macro_dynamics_4}) becomes a time-dependent differential equation at
the thermodynamic limit, $N \rightarrow \infty$.
We substitute $l^m = l$, $l^{m+1} = l+dl$, $R^m=R$ and $R^{m+1}=R+dR$ and simplify, 
and then average Eq. (\ref{R^m_macro_dynamics_4}) over $P(u,v)$ in the same way as for the derivation 
of the differential equation for $l$. Assuming self-averaging for $l$ and $R$, we can rewrite 
Eq. (\ref{R^m_macro_dynamics_4}) as 

\begin{eqnarray}
N(l+dl)(R+dR)&=&NlR+ \langle fv \rangle \nonumber \\
R \frac{dl}{dt} + l \frac{dR}{dt} &=&  \langle fv \rangle .
\label{R^m_macro_dynamics_5}
\end{eqnarray}

\noindent
By substituting Eq. (\ref{l^m_macro_dynamics_3}) into Eq. (\ref{R^m_macro_dynamics_5}) and simplifying, 
we then derive Eq. ( \ref{R^m_macro_dynamics_3}).

\newpage

\begin{table}[p]
\caption{Asymptotic property of the learning curve of the proposed method, Hebbian learning, and perceptron learning.}
\label{asymptotic_property_kappa}
\begin{center}  
\begin{tabular}{c|c}\hline
Learning method&$t$ to power\\ \hline
Hebbian& -0.5\\
perceptron & -0.333 \\
$\kappa=1$ & -0.334 \\
$\kappa=10$ & -0.334 \\
$\kappa=100$& -0.334 \\
$\kappa=1000$ & -0.341 \\ \hline
\end{tabular}
\end{center}
\end{table}

\newpage

%

\begin{figure}[p]
\begin{center}
\includegraphics[width=14cm]{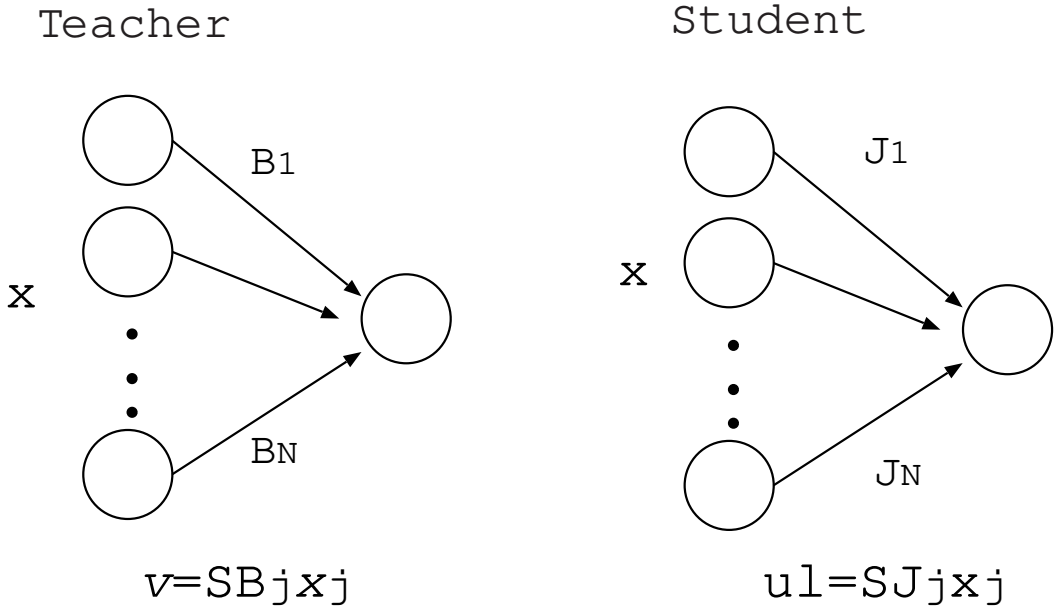}
\end{center}

\caption{Network structure of teacher and student perceptrons.}
\label{network_structure}
\end{figure}

\begin{figure}[p]
\begin{center}
\includegraphics[width=8cm]{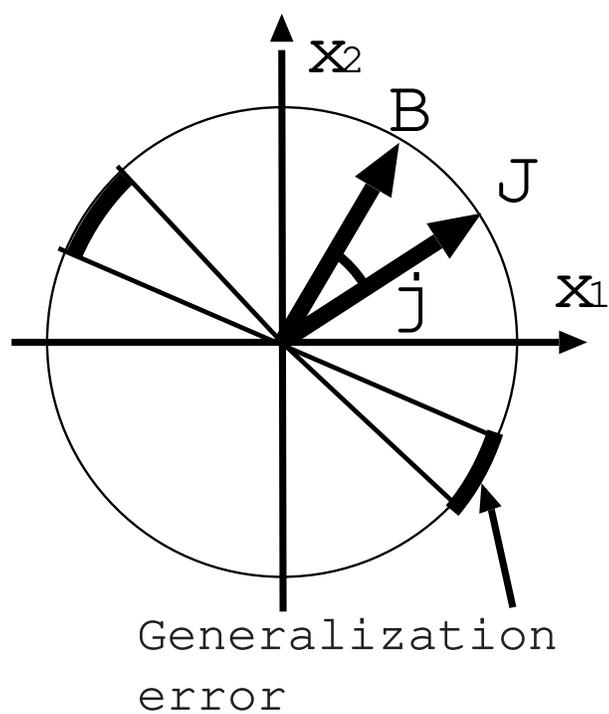}
\end{center}
\caption{Schematic diagram depicting the relationship between the overlap and the generalization error: the weight vector of the teacher and that of the student form an angle of $\varphi$. The teacher output is distinct from the student output on the arcs depicted as thick lines.}
\label{overlap2}
\end{figure}

\begin{figure}[p]
\begin{center}
\includegraphics[width=14cm]{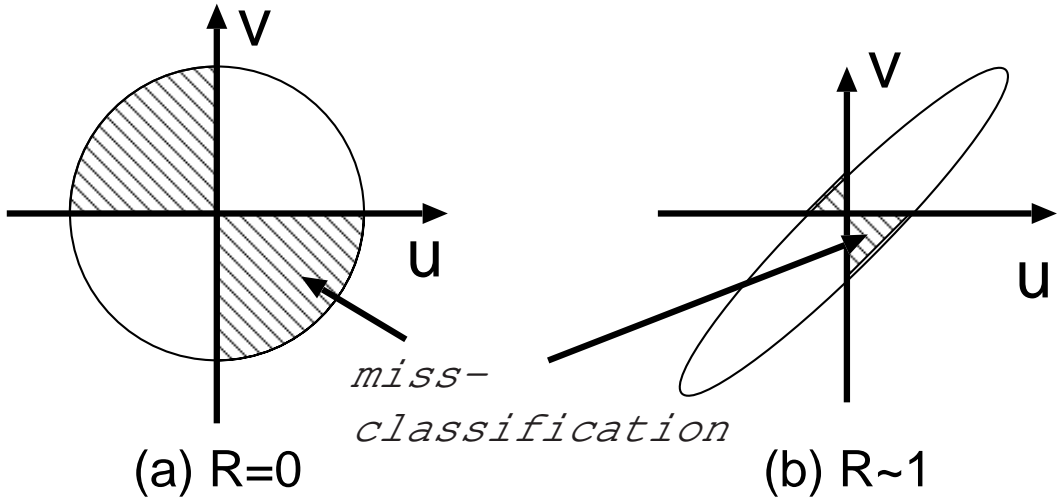}
\end{center}
\caption{Relationship between the overlap and the generalization error in the inner potential space.}
\label{generalization_and_overlap}
\end{figure}

\begin{figure}[p]
\begin{center}
\includegraphics[width=10cm]{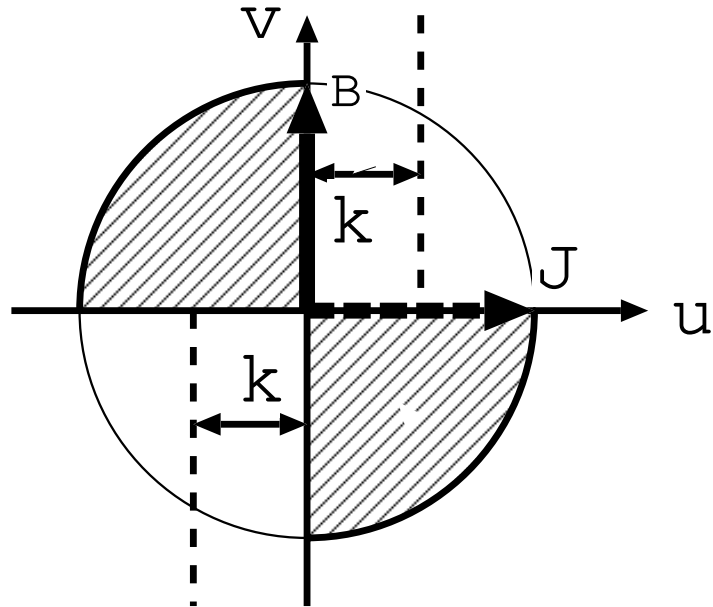}
\end{center}
\caption{Effect of using a margin in the proposed method.}
\label{kappa}
\end{figure}

\newpage

\newpage

\begin{figure}[p]
\begin{minipage}[t]{13cm}
\includegraphics[width=13cm]{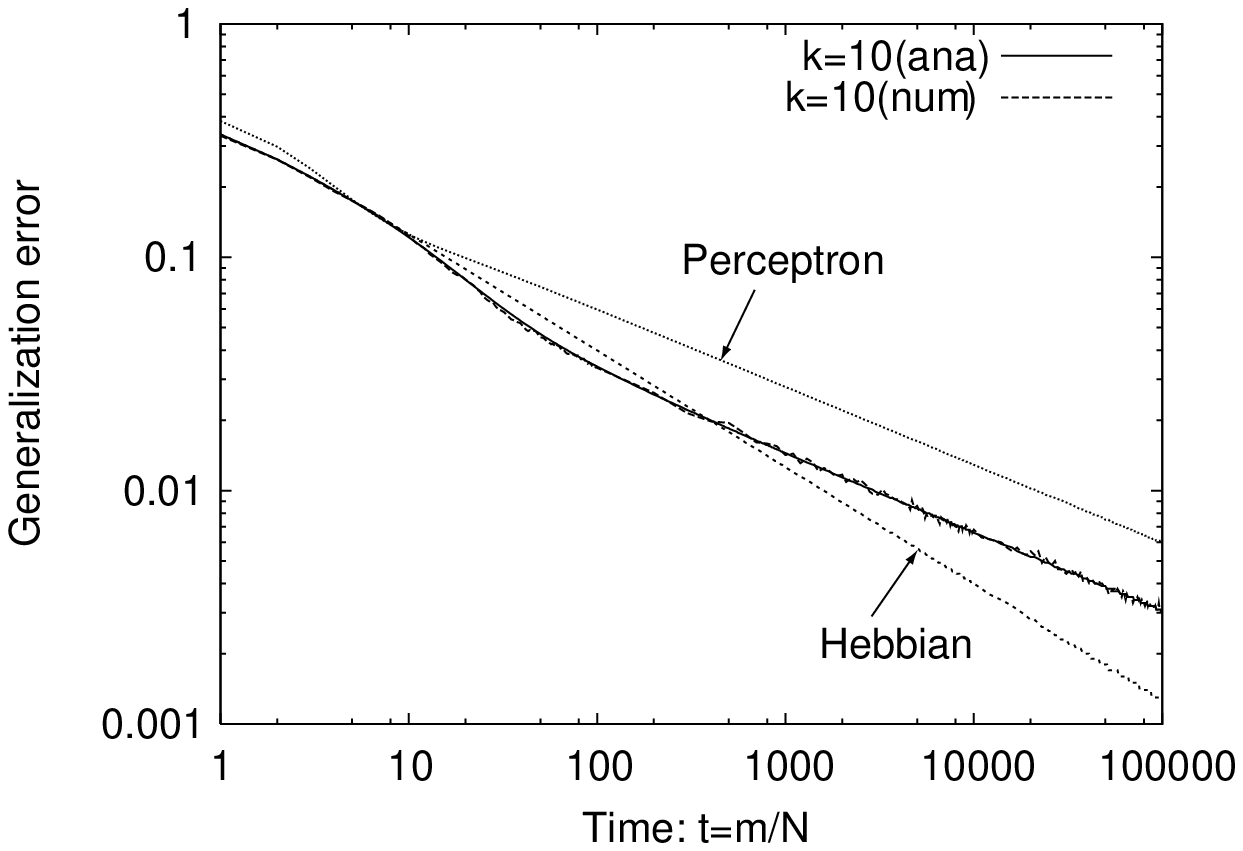}

\begin{center}(a)\end{center}
\end{minipage}
\begin{minipage}[t]{13cm}
\includegraphics[width=13cm]{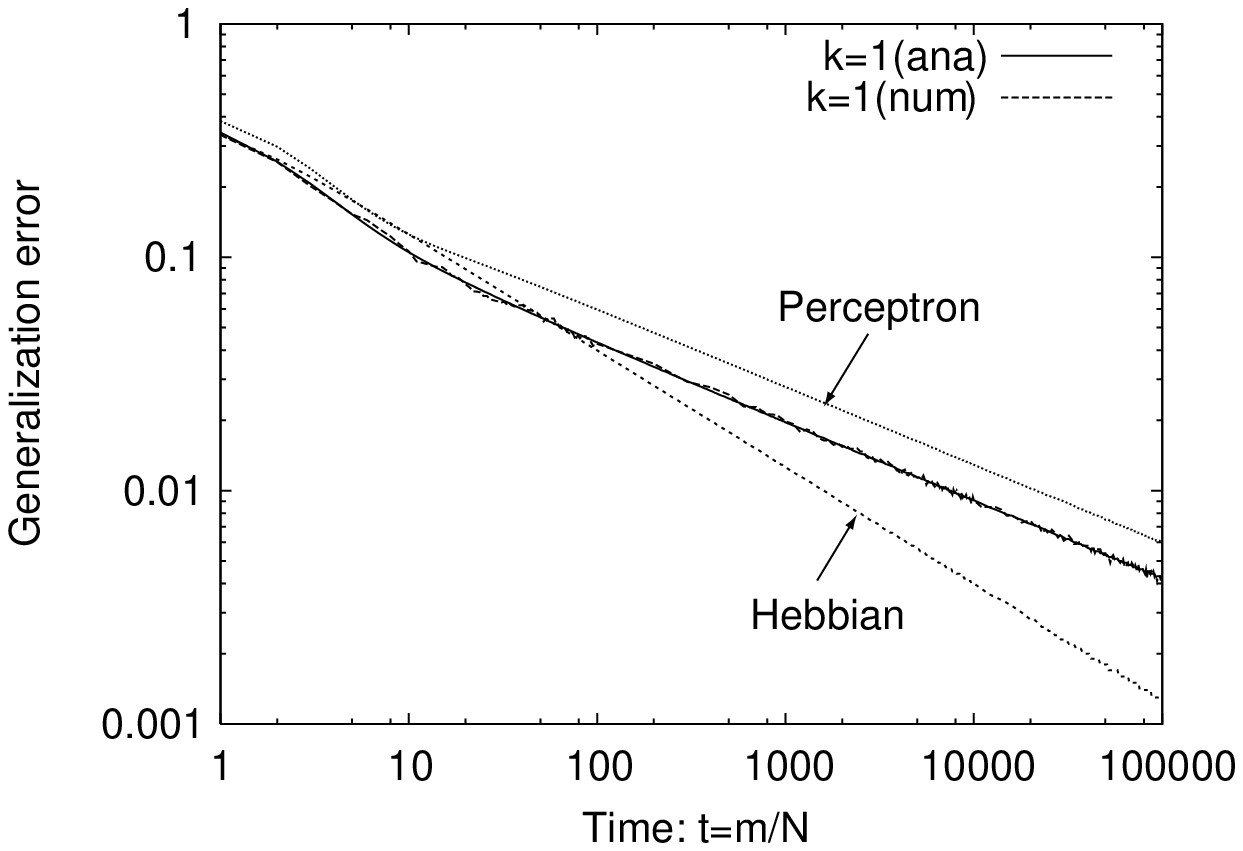}

\begin{center}(b)\end{center}
\end{minipage}
\caption{Learning curves of three learning rules -- the proposed method, 
 Hebbian learning, and perceptron learning -- obtained through analytical solutions. 
The margin $\kappa$ was 10, 1, respectively, for (a) and (b). 
A numerical solution obtained through computer simulations are also shown.}
\label{generalization_error_kappa1}
\end{figure}

\begin{figure}[p]
\begin{minipage}[b]{13cm}
\includegraphics[width=13cm]{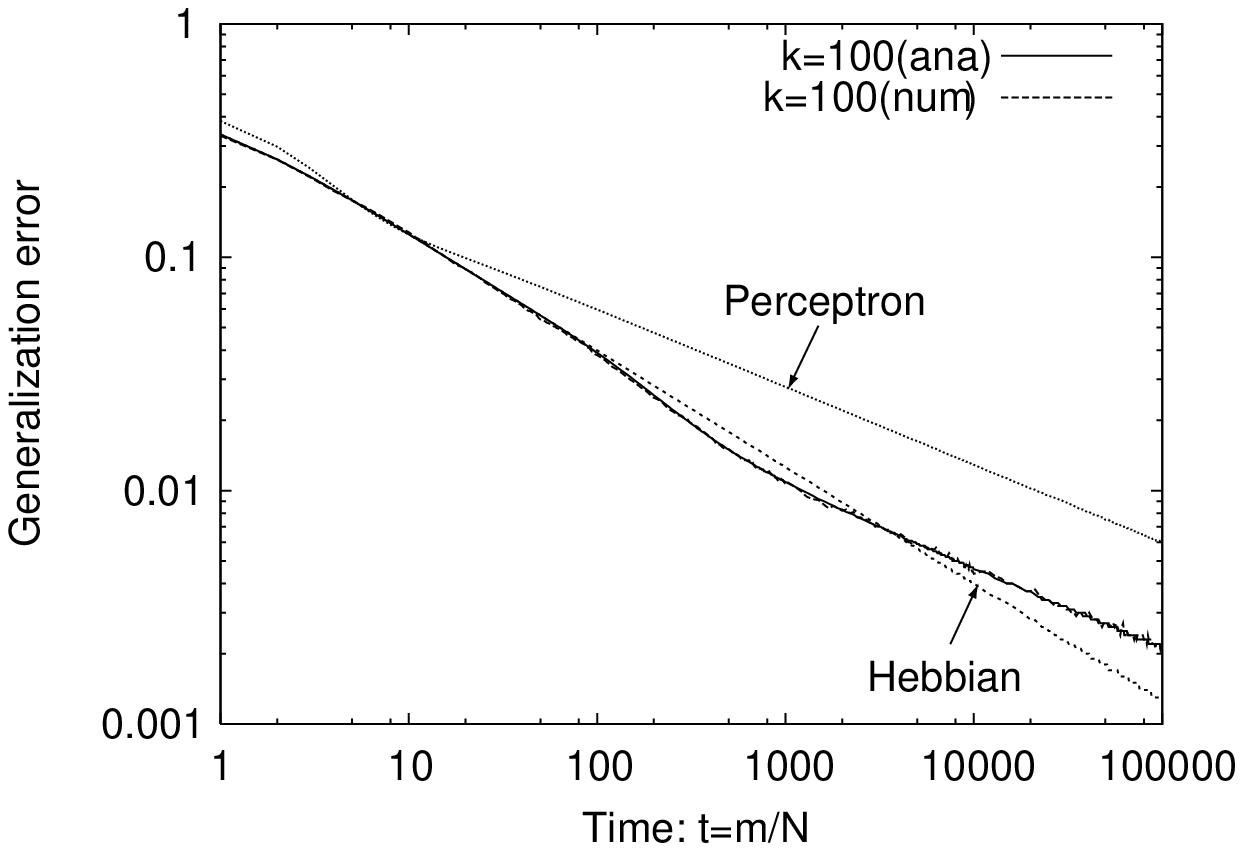}

\begin{center}(a)\end{center}
\end{minipage}
\begin{minipage}[b]{13cm}
\includegraphics[width=13cm]{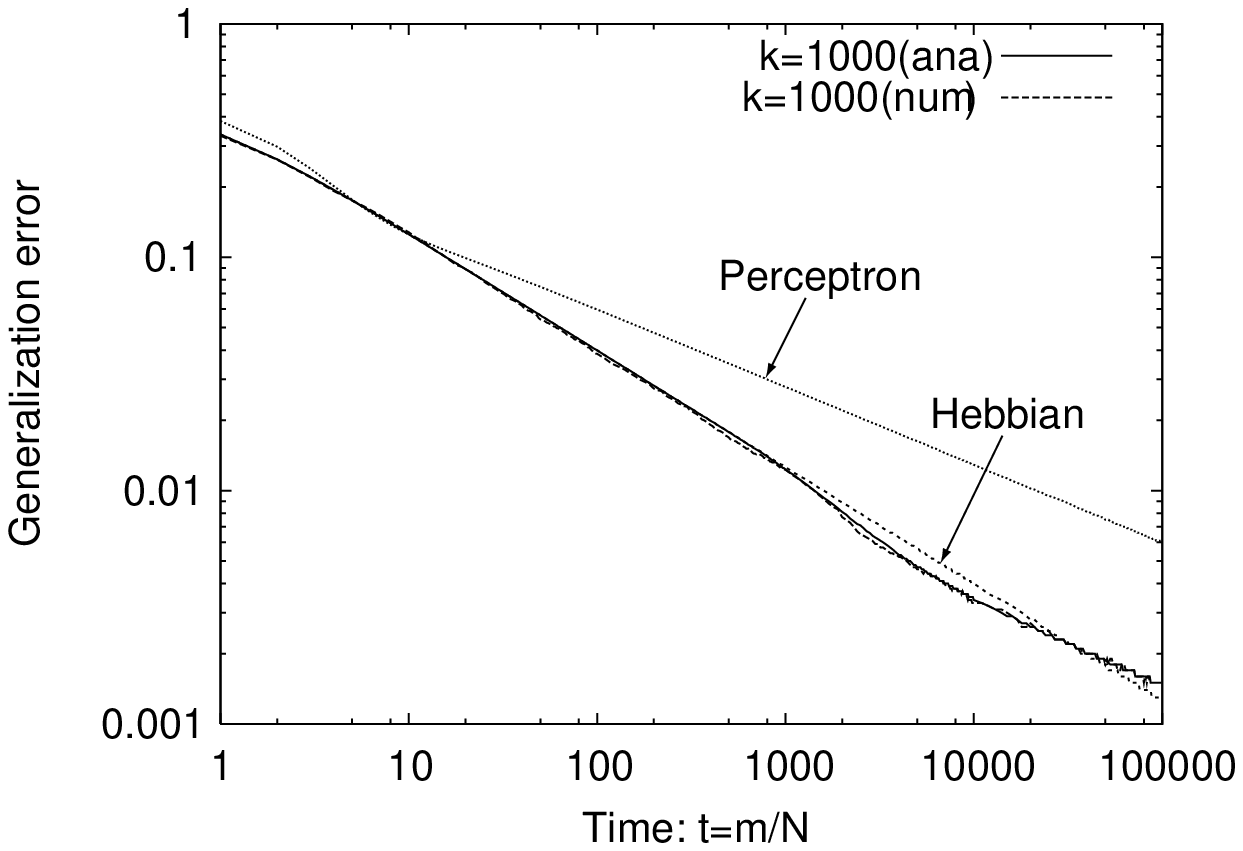}

\begin{center}(b)\end{center}
\end{minipage}

\caption{Learning curves of three learning rules -- the proposed method, 
 Hebbian learning, and perceptron learning -- obtained through analytical solutions. 
The margin $\kappa$ was 100, 1000, respectively, for (a) and (b). 
A numerical solution obtained through computer simulations are also shown.}
\label{generalization_error_kappa2}
\end{figure}

\newpage

\begin{figure}[p]
\begin{minipage}[b]{13cm}
\includegraphics[width=13cm]{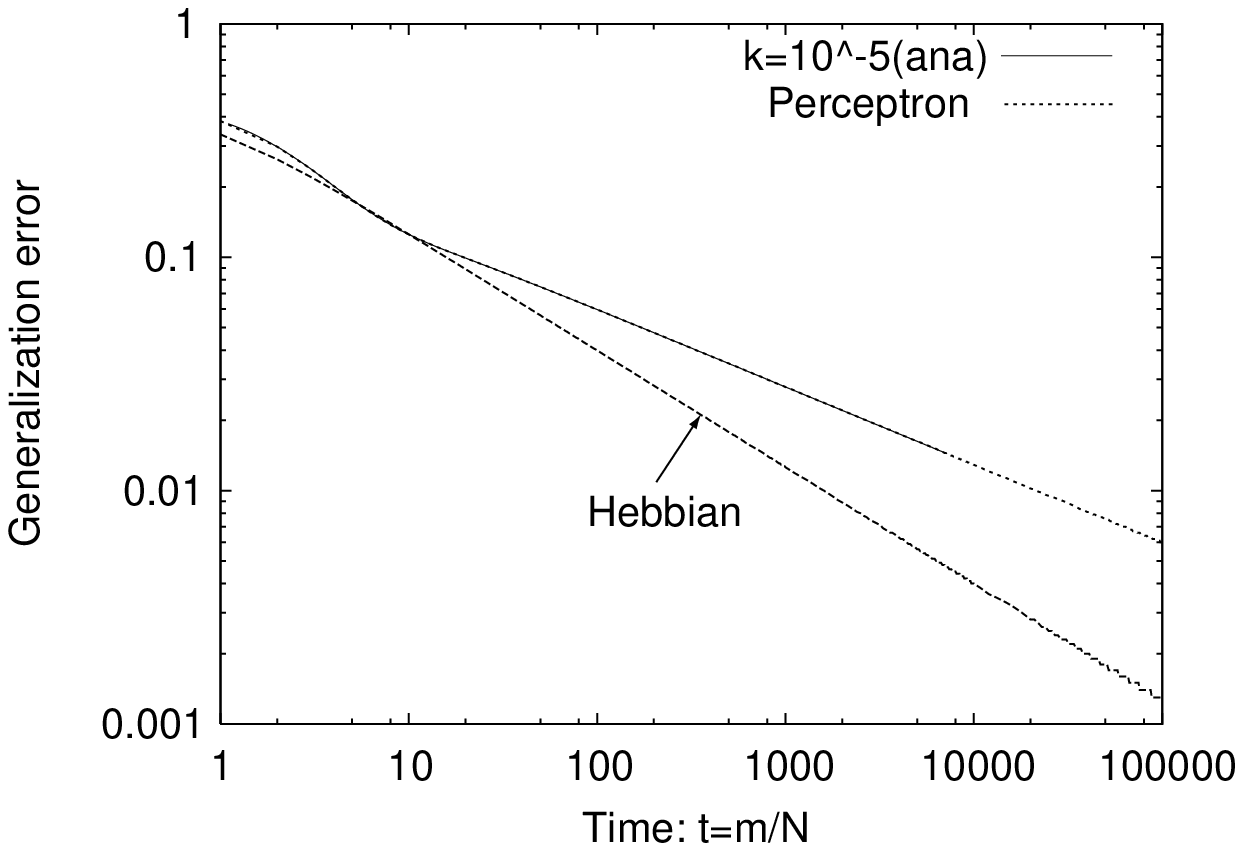}

\begin{center}(a)\end{center}
\end{minipage}
\begin{minipage}[b]{13cm}
\includegraphics[width=13cm]{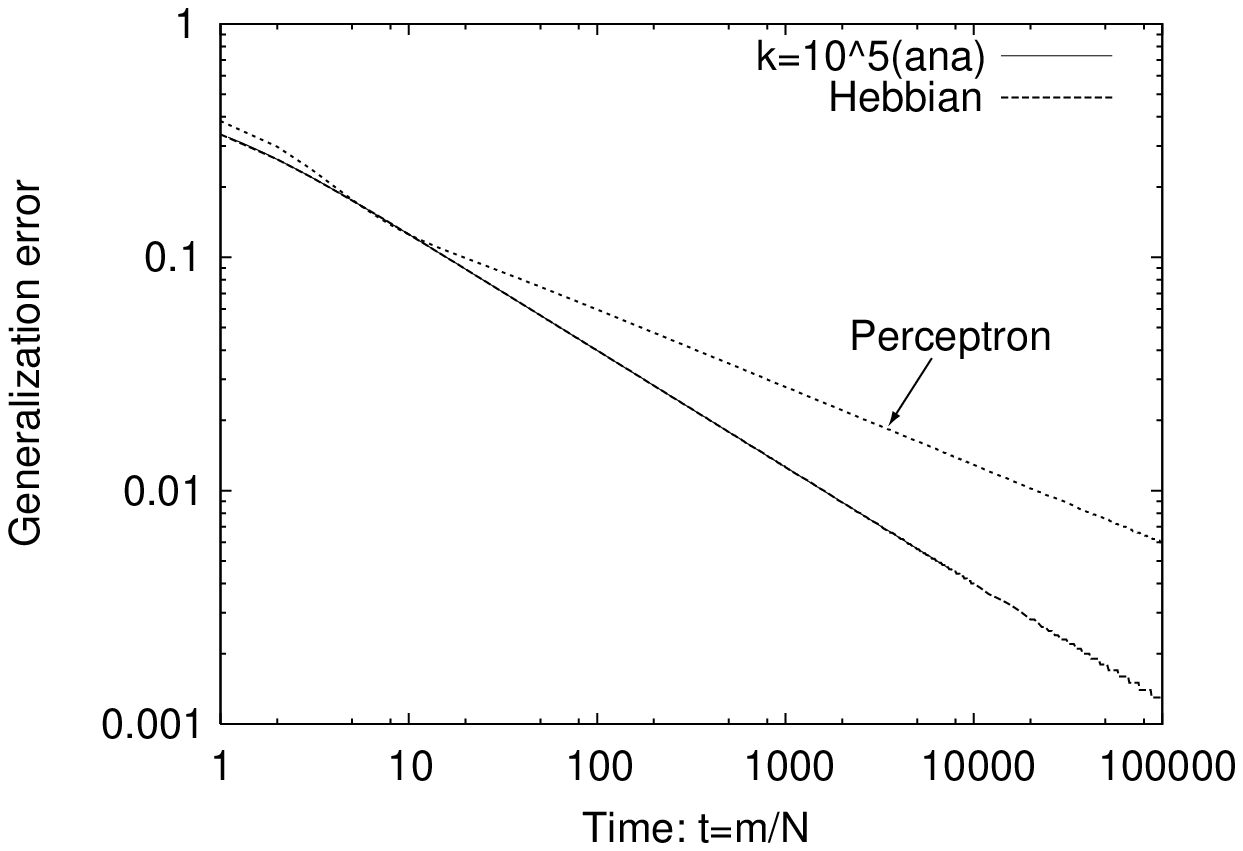}

\begin{center}(b)\end{center}
\end{minipage}
\caption{Learning curves of three learning rules -- the proposed method, 
 Hebbian learning, and perceptron learning -- obtained through analytical solutions. 
The margin $\kappa$ was $10^{-5}$ and $10^5$, respectively, for (a) and (b). 
A numerical solution obtained through computer simulations are also shown.}
\label{generalization_error_kappa3}
\end{figure}

\newpage

\begin{figure}[p]

\begin{minipage}[t]{13cm}
\includegraphics[width=13cm]{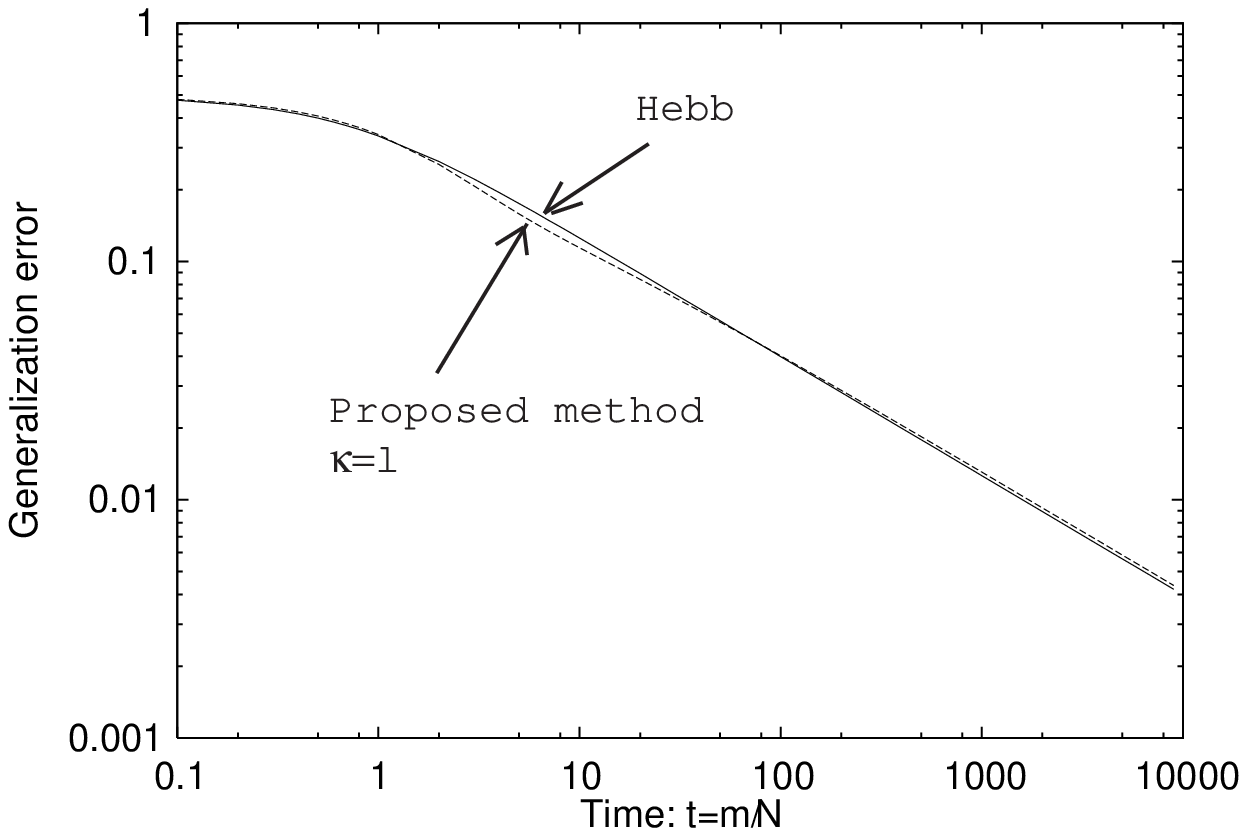}

\begin{center}
(a) $\kappa=l$
\end{center}
\end{minipage}

\begin{minipage}[t]{13cm}
\includegraphics[width=13cm]{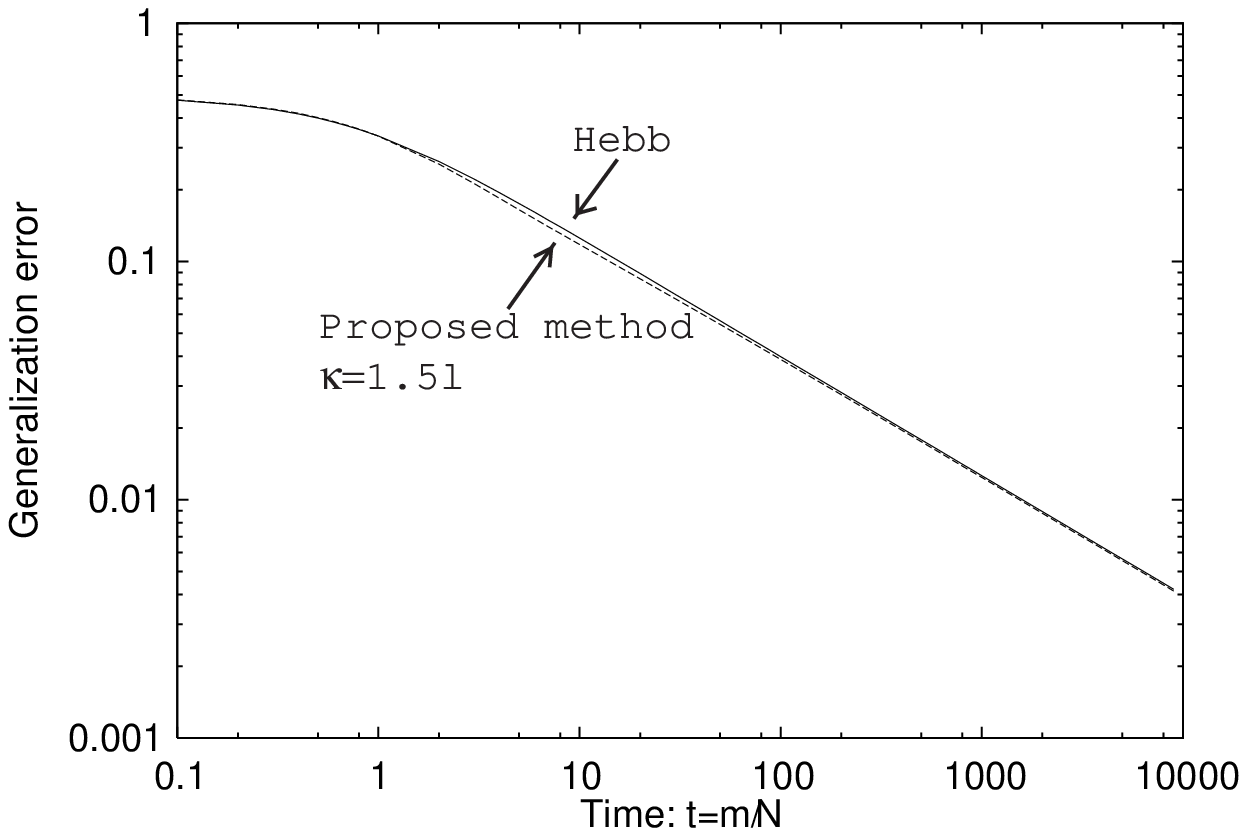}

\begin{center}
(b) $\kappa=1.5l$
\end{center}
\end{minipage}
\caption{Dynamics of the generalization error with an adaptively controlled margin. (a) $\kappa=l$, (2) $\kappa=1.5l$}
\label{Adaptive_margin1}
\end{figure}

\newpage

\begin{figure}
\begin{minipage}[t]{13cm}
\includegraphics[width=13cm]{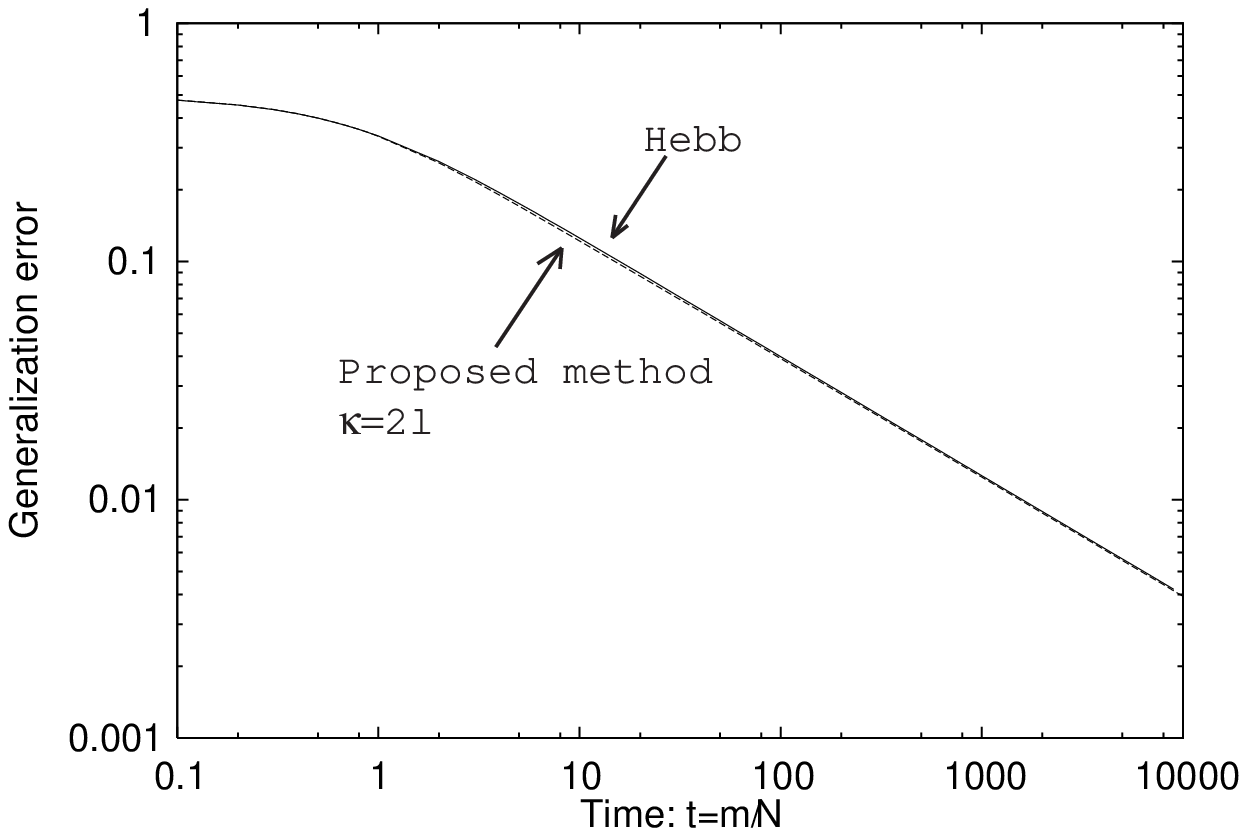}

\begin{center}
(c) $\kappa=2l$
\end{center}
\end{minipage}
\caption{Dynamics of the generalization error with an adaptively controlled margin. $\kappa=2l$}
\label{Adaptive_margin2}
\end{figure}

\newpage

\begin{figure}[p]
\noindent
{\bf \large Figure Legends}

Figure 1:Network structure of teacher and student perceptrons.

Figure 2:Schematic diagram depicting the relationship between the overlap and the generalization error: the weight vector of the teacher and that of the student form an angle of $\varphi$. The teacher output is distinct from the student output on the arcs depicted as thick lines.

Figure 3:Relationship between the overlap and the generalization error in the inner potential space.

Figure 4:Effect of using a margin in the proposed algorithm.

Figure 5:Learning curves of three learning rules -- the proposed algorithm,  Hebbian learning, and perceptron learning -- obtained through analytical solutions. The margin $\kappa$ was 10 and 1, respectively, for (a), (b). A numerical solutions obtained through computer simulation are also shown.

Figure 6:Learning curves of three learning rules -- the proposed algorithm,  Hebbian learning, and perceptron learning -- obtained through analytical solutions. The margin $\kappa$ was 100 and 1000, respectively, for (a), (b). A numerical solutions obtained through computer simulation are also shown.

Figure 7:Learning curves of three learning rules -- the proposed method, 
 Hebbian learning, and perceptron learning -- obtained through analytical solutions. 
The margin $\kappa$ was $10^{-5}$ and $10^5$, respectively, for (a) and (b). 
A numerical solution obtained through computer simulations are also shown.

Figure 8:Dynamics of the generalization error with an adaptively controlled margin. $\kappa=l$ and $\kappa=1.5l$.

Figure 9:Dynamics of the generalization error with an adaptively controlled margin. $\kappa=2l$.
\end{figure}


\begin{thebibliography}{99}
\bibitem{Nishimori1999} 
Nishimori H., (1999). 
Spin Glass Theory and Statistical Mechanics of Information.:
Iwanami shoten. 
\bibitem{Vallet1989}
Vallet F. (1989). 
Europhys. Lett. {\bf 9}, 315.
\bibitem{Biehl1992}
Biehl M. and Schwarze H., (1992).  
Europhys.  Lett. {\bf 20}, 733.
\bibitem{Gardner1988}
Gardner, E., (1988).  
J. Phys. A: Math. Gen. {\bf 21}, 257.
\bibitem{Rosenblatt1962}
Rosenblatt, F., (1962). 
Principles of Neurodynamics.:
Spartan.
\bibitem{Hertz1991}
Hertz J. Krogh A. and Palmer R. G., (1991). 
Introduction to the Theory of Neural Computation.:
Addison Wesley.
\bibitem{Kinouchi1992}
Kinouchi O. and Caticha N., (1992).
J. Phys. A: Math. Gen. {\bf 26}, 6243.
\end{thebibliography}
\end{document}